# Photo-memristive sensing with charge storing 2D carbon nitrides

A. Gouder[1,2], A. Jiménez-Solano[1], N. M. Vargas-Barbosa[4], F. Podjaski[1,3]*, B. V. Lotsch[1,2,3]*

**Affiliations**

[1]: Max Planck Institute for Solid State Research, Heisenbergstraße 1, 70569 Stuttgart, Germany

[2]: Department Chemistry, Ludwig-Maximilians-University Munich, Butenandtstraße 5-13, 81377 Munich, Germany

[3]: Cluster of excellence e-conversion, Schellingstraße 4, 80799 Munich, Germany

[4]: Institute for Energy and Climate Research (IEK-12): Helmholtz Institute Münster, Forschungszentrum Jülich, Corrensstraße 46, 48148 Münster, Germany

* Correspondence to: b.lotsch@fkf.mpg.de, f.podjaski@fkf.mpg.de

**Abstract**

We report the charge storing 2D carbon nitride potassium poly(heptazine imide), K-PHI, as a direct memristive (bio)sensing platform. Memristive devices have the potential to innovate current (bio)electronic systems such as photo-electrochemical sensors by incorporating new sensing capabilities including non-invasive, wireless remote and time-delayed (memory) readout. We demonstrate a direct photomemristive sensing platform that capitalizes on K-PHI's visible light bandgap, large oxidation potential and intrinsic optoionic light energy storage properties. Our system simultaneously enables analyte concentration information storage as well as potentiometric, impedimetric and coulometric readouts on the same material, with no additional reagents required. Utilizing the light-induced charge storage function of K-PHI, we demonstrate analyte sensing *via* charge accumulation and present various methods to write/erase this information from the material. Additionally, fully wireless colorimetric and fluorometric detection of the charged state of K-PHI is demonstrated and could facilitate its use as particle-based *in-situ* sensing probe. The various readout options of the K-PHI's response enable us to adapt the sensitivities and dynamic ranges without modifying the sensor. We demonstrate these features using glucose as an example analyte over a wide range of concentrations (50 µM to 50 mM). Moreover, due to the strong oxidative power of K-PHI, this sensing platform is able to detect a large variety of organic or biologically relevant analytes. Since PHI is easily synthesized, based on earth abundant precursors, biocompatible, chemically robust and responsive to visible light, we anticipate that the sensing platform presented herein opens up novel memristive and neuromorphic functions.

**Introduction**

The trend towards digitization and automation necessitates novel sensing and information storage concepts. Biosensors are of critical importance for medical applications, in smart healthcare systems and environmental monitoring.[1] One of the most prominent fields of biosensing is glucose monitoring due to its high and practical value in diabetes management.[2] Most modern 3rd generation electrochemical sensors incorporate the enzyme glucose oxidase (GOx) as the receptor due to its very good selectivity,[3] but suffer from drawbacks, such as high price, limited reproducibility, lack of stability and complex enzyme immobilization processing steps.[4] Nanozyme biosensors utilize nanomaterials to replace GOx, but suffer instead from low selectivity, operate far from physiological conditions (e.g., alkaline pH) and thus far have not reached any practical application.[2] Alternative sensor designs based on biocompatible yet robust materials are therefore vital for the future growth of this field.

Emerging neuromorphic sensing and monitoring concepts utilize memristive effects to tackle limitations of current devices.[5–8] The traditional path to application of memristive devices are non-volatile memory applications with many distinctly addressable states.[5] Artificial synapses are a class of bio-memristive devices which are cheap and offer biocompatibility.[5,9,10] However, memristors have further potential. Coupling photovoltaics with memristors or transistors can drastically reduce computational energy consumption with photonic- or photomemristive devices, which have been dubbed as "solar brains".[11] The fast growing field of memristive biosensors addresses several limitations of the current state-of-the-art sensors such as ultra-high sensitivity down to the fM concentration range for biological materials.[7,12] Most current chemical memristive sensors are comprised of classical inorganic, memristive materials (e.g. Si-nanowires) which are functionalized with a receptor and produce a voltage gap in their memristive behavior, which can be linked to the analyte concentration.[7] However, such two-terminal memristor devices suffer from complex device geometries, as they do not have a straightforward pathway for biomolecules to approach the device.[13] To address this challenge, complex surface patterning[14] or nanowire[15] designs are necessary. While most memristive sensors so far have focused on the detection of complex biomolecules such as DNA aptamers,[16] cancer markers[17] or the ebola and dengue virus,[18,19] sensing of easier biomolecules such as glucose was demonstrated but with a lack in both sensitivity (10-40 mM) and selectivity.[20]

Organic memristive devices have been designed with several different voltage thresholds or volatile switching mechanisms.[5,6,21–23] The common denominator is the writing and reading of different conductance states, which can be utilized for information storage. Redox-based switching uses electrochemical redox reactions of polymers (e.g. PEDOT:PSS)[5,23] accompanied by ion diffusion, which modifies the conductance of the material.

Carbon nitrides are an interesting class of organic of materials with potential application in biosensing and memristive devices. These are layered molecular materials with one- (1D) or two-dimensional (2D) triazine- or heptazine backbones and a visible-light bandgap. Carbon nitrides have lately attracted interest in various research areas including photocatalysis,[24–27] electrochemical and solar energy storage,[24,28,29] molecular machines,[30] environmental remediation,[31] and non-enzymatic or 'nanozyme' sensing of e.g. glucose.[32–34] Potassium polyheptazine imide (K-PHI), a recently discovered 2D carbon nitride,[26,35,36] has remarkable optoionic properties[37] that are derived from its dual functionality of light harvesting and charge storage.[24,28] This versatile toolkit has led to the design of novel responsive and/or charge storing devices, such as a solar battery anode with a capacity of 12.1 mAh/g.[28] K-PHI's large bandgap of ~2.7 eV (see Fig. S1), corresponding to an absorption edge in the blue region of the visible spectrum, together with the material's suitably positioned band edges (+2.2 and -0.5 V vs. NHE, respectively), provides a large thermodynamic driving force for oxidation and reduction of various chemical species,[38] while being chemically robust. Charge storage is accompanied with a change in the material's optical and electrochemical properties.[28,37] A color change from yellow to blue upon photoreduction is the most prominent of them and a signature of free electrons, translating into photodoping.[24,26,28,39,40]

Herein, we utilize these intrinsic memristive material property changes of K-PHI that result from environmental interactions to propose a refined photomemristive sensor design, instead of the commonly used two-terminal design. We show that K-PHI acts as the receptor unit and transducer coupled to a memristive amplifier, thereby combining all sensing components in the same material (direct sensing) and propose various measurement techniques as readout methods. We first show the direct photo-electrochemical (PEC) sensing ability of K-PHI (Fig. 1a, left) with different analytes and then demonstrate the K-PHI memsensing concept via different electrochemical readout schemes (potentiometric, impedimetric and coulometric; Fig. 1a, right). Further, we rationalize the material's applicability as wireless, memristive sensor via optical readout methods (colorimetric and fluorometric), using glucose sensing as a case study for the photomemristive sensing. Thus, we connect the field of memristive sensors to the well-established electrochemical sensor design by moving away from the more complex "classical" memristor design to a direct PEC sensor with built-in memristive functionalities. This is only possible due to the special photo-electrochemical properties of the novel carbon nitride K-PHI, namely light absorption, charge storage and a large oxidative driving force.

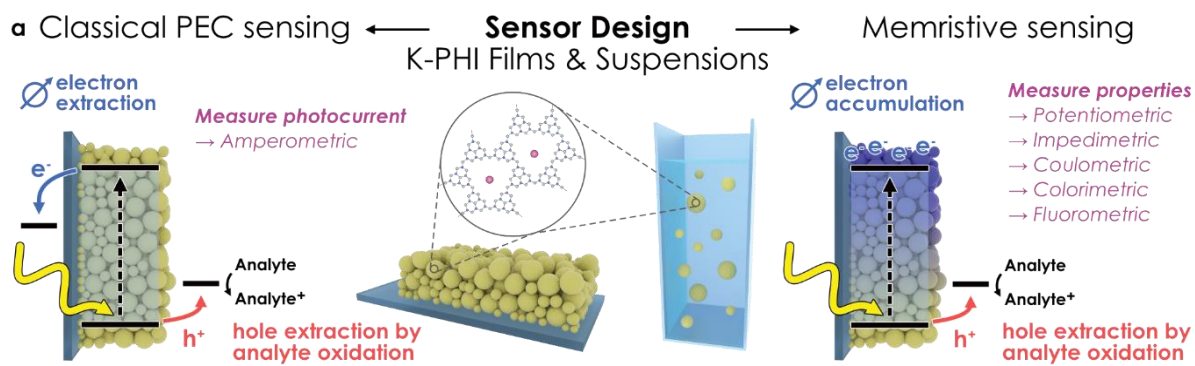

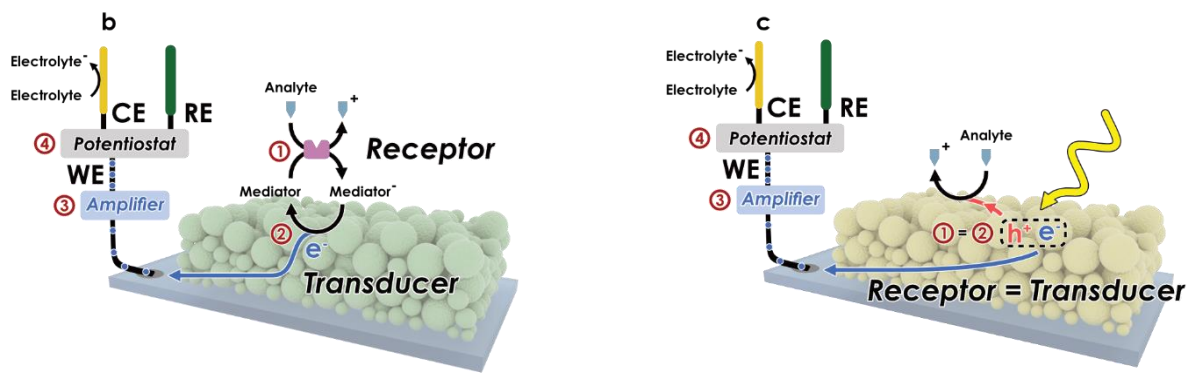

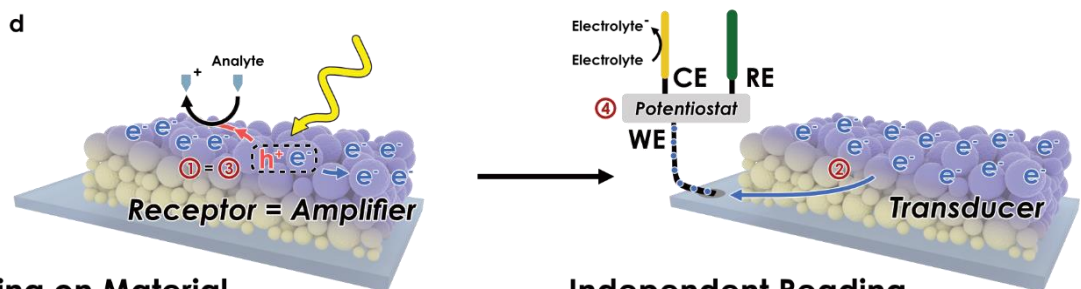

**Fig. 1 | Concept of an oxidative (memristive) photo-electrochemical sensor. a,** The carbon nitride K-PHI can be deployed as a thin film on a conductive substrate or as wireless particles in an aqueous suspension (middle). When used as PEC sensor, photoexcited charges generate current that is usually mirrored by holes oxidizing the analyte. The electrons are extracted continuously, reflecting the reaction rate (left). On the other hand, a memristive sensor as proposed herein accumulates the electrons and stores the concentration information in the material, leading to a modified material state (right). The change in material properties can be analyzed and quantified via different electrochemical and optical methods. **b-c,** Components of carbon nitride-based indirect electrochemical (b) or direct photo-electrochemical (c) sensors.[41,42] In the first (indirect) case, a receptor ① generates a chemical signal proportional to the analyte concentration. The chemical signal is transferred *via* a mediator to the physicochemical transducer ②, which transforms the chemical to an electric signal.

This signal is amplified ③ and then detected with a potentiostat ④ in an electrochemical cell consisting of the sensor (working electrode, WE), a suitable counter (CE) and reference electrode (RE). Note that in this work we do not explore this approach. In the second case (c), electron-hole pairs are generated upon illumination. The holes are extracted *via* oxidation of the analyte. The electrons generate a photocurrent proportional to the analyte concentration. As the receptor is the same material as the transducer ① = ②, we denote it as a "direct" sensor. **d,** In a photomemristive sensor, photogenerated holes oxidize the analyte ① analogous to the direct amperometric sensor. However, instead of generating a photocurrent directly, the electrons charge the sensor (writing) and thus, change the sensor's material properties as a function of the analyte concentration. As this change is integrative due to accumulation of charge, the sensor also acts as an amplifier ③ while being adaptable to measurement conditions. Subsequently, the material properties can be accessed via electrochemical or optical techniques (readout) ②, ④.

**Direct PEC sensing with K-PHI**

The concept of photo-electrochemical (PEC) biosensors has been investigated in great detail and briefly discussed above (Fig. 1b, c).[43] To describe the novel direct memristive sensing mode (see next section), we first investigate the capability of K-PHI films (see Fig S2 and S3 for details) to perform direct PEC amperometric sensing, i.e., to simultaneously act as receptor and transducer while being illuminated in aqueous conditions (Fig. 1c). We have investigated glucose sensing as a case study (see supporting information (SI) Section 2), relating the analyte concentration to the photocurrent. Two linear concentration ranges are observed (0-1 mM and 1-10 mM) and a limit of detection (LOD) of 11.4 µM (0.21 mg/dL) can be determined from the lower concentration range.[44] These observations are comparable to reported linear ranges and LOD's for metal-free amperometric carbon nitride glucose sensors (LOD of 11 µM and a linear range from 1 to 12 mM),[34] which additionally require the mediator $H_2O_2$ in convolution with the enzyme GOx as the receptor. The much simpler, single-component design presented here underscores the potential of K-PHI as a direct, non-enzymatic amperometric PEC glucose sensor. A sensitivity increase of 1-2 orders of magnitude was reported for more complex systems utilizing heterojunctions with inorganic materials tailored for glucose detection, yet at the expense of analyte flexibility and biocompatibility.[45,46]

To highlight the broad application range of K-PHI sensors, we have analyzed other sugars, alcohols, as well as typical electron donor molecules, which are widely used as sacrificial agents in photocatalytic applications, at a concentration of 1 and 5 mM (Fig. 2a). All these analytes show a strong photocurrent response and hence, can be detected efficiently with K-PHI sensors. At a concentration of 5 mM, the mono- and disaccharide sugars (Fig. 2a, red) produce photocurrents between ~0.6 and 1 µA/cm$^2$. By far the strongest response is observed for triethanolamine (TEOA) and 4-methylbenzyl alcohol (4-MBA) - both typical sacrificial electron donors for photocatalytic studies[24,25,35,47] - with photocurrents of 7.5 and 23 µA/cm². respectively. While it is unclear whether the origin of these large photocurrents is thermodynamic or kinetic in nature, or both, it suggests that these donors can bypass possible kinetic bottlenecks on the hole extraction step. We note that while the evaluation of sacrificial electron donors in photocatalysis is still largely phenomenological and highly convoluted with the electron extraction rate, the above measurement may be used to isolate and quantify donor reaction rates, i.e., hole quenching efficiencies.[36,47–49] The physiologically relevant molecules ascorbic acid (AA) and uric acid (UA) were measured at lower concentrations of 0.1 and 0.5 mM, as usually found in blood serum.[50] A significant photocurrent response of 0.38 and 0.35 µA/cm$^2$ is observed already at a concentration of 0.1 mM for AA and UA, respectively, suggesting a viable monitoring strategy for these analytes as well. However, the strong oxidative driving force, which enables considerable photocurrents with many different species, makes it difficult to differentiate between organic substances in mixtures. This selectivity limitation could be addressed by suitable functionalization of the K-PHI.[49,51]

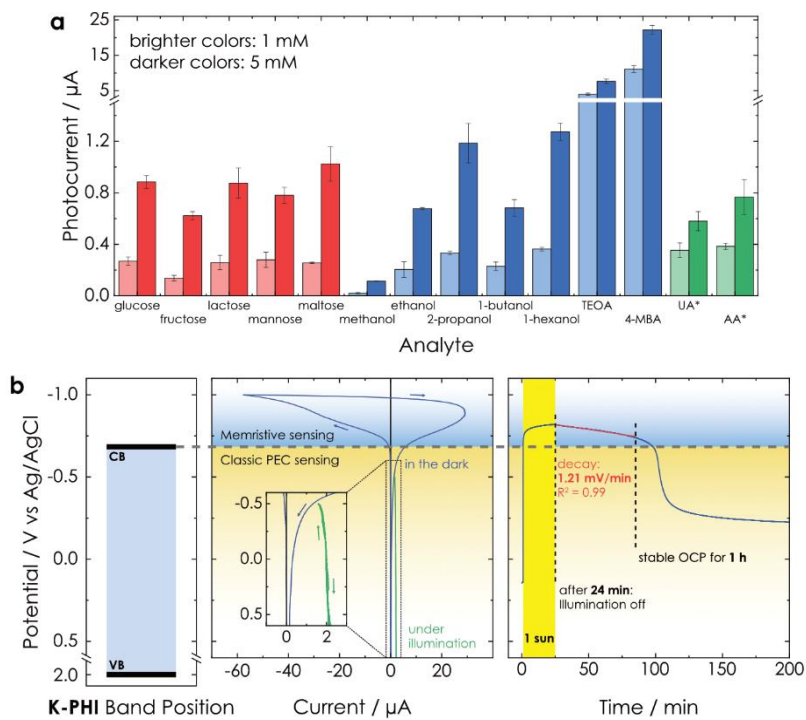

**Fig. 2 | (Photo)electrochemical properties of K-PHI for direct and memristive sensing. a,** Photocurrent of a direct amperometric K-PHI sensor for different sugars (red), alcohols (blue) and the physiologically relevant molecules ascorbic acid (AA) and uric acid (UA) (green). Measurements are performed at a concentration of 1 and 5 mM for sugars and alcohols, and 0.1 and 0.5 mM for UA and AA with a bias of 0 V vs. Ag/AgCl, highlighting the versatility of K-PHI sensors. **b,** Left: Valence band (VB) and conduction band (CB) positions referenced to the Ag/AgCl reference electrode potential. Middle: Cyclic voltammetry measurement at 50 mV/s, showing the behavior of K-PHI in the relevant potential windows for sensing. At potentials more positive than -0.7 V vs. Ag/AgCl, a constant photocurrent is observed (green), which is used for direct PEC sensing. Charge storage occurs at potentials more negative than -0.7 V vs. Ag/AgCl. Right: Open circuit potential (OCP) evolution of K-PHI in presence of glucose during illumination (photocharging) and its subsequent stability in the dark.

**Photo-Memristive direct sensing with electrochemical readout**

Next, we demonstrate the memristive operation of the K-PHI sensor (Fig. 1d and Fig. 3a). The necessary double functionality of charge storage and light absorption is illustrated in Fig. 1b (middle). Electric charging via cyclic voltammetry (CV) measurement shows the charging and subsequent discharging of the material negative of -0.7 V vs. Ag/AgCl, where the material's highly capacitive regime starts. Upon illumination with 1 Sun (AM 1.5 G) at open circuit potential (OCP) and in the presence of an oxidizable analyte, a photopotential immediately develops by electron accumulation in this potential range, which remains stable for ~1 h after 20 min of illumination. The photopotential decay rate is very slow (1.21 mV/min) and can be attributed to self-discharge via the FTO substrate promoting hydrogen evolution, since charge storage occurs at potentials more negative than the hydrogen reduction potential (ca. -0.61 V vs. Ag/AgCl at pH 7).[28] The inherent charge storage properties of K-PHI enable a quantitative comparison between the analyte concentration and the charging state of the material when the illumination time is fixed. The photomemristive behavior integrates the analyte oxidation current and produces the stable change in the optoelectronic properties of the K-PHI as the measurand. This accumulated charge is proportional to the charging current, which depends on both analyte concentration and illumination time. Thus, a longer illumination (or writing cycle) can improve the sensitivity for low analyte concentrations, as the system has more time to interact with the analyte and accumulate charges. In principle, there is no lower analyte detection limit in the absence of self-discharge (and other experimental limitations such as parasitic oxidation reactions due to a lack of selectivity), as the illumination time can be increased at will. On the other hand, a short light pulse allows for fast sensing of high analyte concentrations. Thus, K-PHI can act as receptor and transducer (as discussed in the previous paragraph) but also as a memristive amplifier, hence combining all sensor components on a single material. In the following, three electrochemical readout methods of the sensor state are demonstrated and explained (Fig 3a).

*Potentiometric sensing* at open circuit condition (OCP) evaluates the photopotential that a given K-PHI thin film reaches after illumination with a fixed duration and intensity (1 Sun), in the presence of a specified analyte concentration (here glucose). After activation of the sample (SI Section 3), upon illumination the OCP shifts negatively over time (Fig. 3b), which shows that the sensor is being charged. With increasing analyte concentration, the OCP response becomes more pronounced (shifts faster and to more negative potentials), in line with increasing oxidation currents shown in direct PEC measurements (Fig. S4). In all cases, we observe a nearly linear regime until ap-prox. -0.7 V vs. Ag/AgCl, followed by a second and significantly slower saturation regime. The gradient of the OCP shift depends on the differential charge density or capacity of K-PHI (see CV in Fig. 1b), which is comparably small before reaching -0.7 V vs. Ag/AgCl and primarily due to surface capacitance,[28] followed by a faradaic response with a higher differential capacity. The different charging kinetics can be exploited by using different illumination times to sense different effective analyte concentration ranges, thereby adapting the material's response window and hence, relative sensitivity. This becomes more evident when looking at the OCP evolution with increasing illumination times (Fig. 3c): At short illumination times of 1-4 seconds (red shading in Fig. 3b and c), higher glucose concentrations (10-50 mM) yield the most significant potential change. They can be measured more precisely as the OCP does not reach a value more negative than -0.7 V vs. Ag/AgCl, i.e., no saturation occurs yet. When illuminating longer than 4 seconds (blue area in Fig. 3b and c), more charges accumulate over time at every concentration, which induces a stronger potential change. Hence, a steeper slope for small concentrations of 1-5 mM can be seen in Fig. 3c, which enables more accurate sensing of low analyte amounts due to the extended interaction and memristive integration times. The illumination time can be used as a parameter to adapt the sensitivity of this potentiometric sensor (see Table 1). A calibration trial is necessary to find such suitable illumination times for the respective concentration ranges and sensitivities. In principle, similar adjustments are possible by changing the light intensity and hence, photon flux, which generates the photoresponse. To extract the sensing information, a relationship between the sensing signal (photopotential) and analyte concentration is required. We have performed both phenomenological non-linear fitting and linear fitting of the sensor's response. The non-linear fitting (see SI Section 5.1) uses an equation that mimics the contributions of the charging mechanism including their saturation, without requiring dedicated input parameters for a given set of conditions. With this non-linear approach, the response of all electrochemical and optical readout modes over the entire concentration range is possible, with a much better fitting quality compared to the linear fit. A detailed description of the fitting is given in SI Section 5.1. The linear ranges are given in Table S1 and the fit is shown in Fig. S9. The non-linear fit is shown in Fig. 3c, Fig. S7 and Fig. S8.

*Impedimetric sensing* utilizes electrochemical impedance spectroscopy (EIS) and probes the change in resistance of the charged state, which is equivalent to a direct readout of the memristive state. The measurements are performed after illumination with 1 Sun for 30 s at OCP by a 10 mV AC perturbation signal to extract the sensor's resistance.

When increasing the analyte concentration and hence, charge accumulation on the K-PHI analogous to the potentiometric sensing discussed above (Fig. 3b), a decrease in magnitude of the impedance is measured and shown via a Bode plot in Fig. 3d. We interpret this as the material becoming more conductive (less resistive) in response to charge accumulation of mobile charge carriers (photogenerated electrons and intercalated ions interacting with the backbone, equivalent to photodoping),[28,37,40] as a consequence of the interaction with the analyte. The resistance can be interpreted as direct measure of the memristance $R_M$ of the system as it directly and monotonously relates to the charged state. The systematic relationship between the magnitude of the impedance and analyte concentration at different frequencies is shown in Fig. 3e. In the frequency region < 100 Hz, a change in impedance is observed that contains the sensing information. At frequencies < 1 Hz (red shading in Fig. 3d and e), the most pronounced shift with respect to the analyte concentration is observed, which causes a steeper slope for both small and large concentrations. We attribute this low frequency range to the comparably slow faradaic charge storage process being triggered. It is kinetically slower than the double layer capacitance and therefore more visible at lower frequencies. The advantage of measuring at moderately higher frequencies (1-100 Hz) is the shorter measurement time (blue shading in Fig. 3d and e). However, for large analyte concentrations the difference in magnitude of impedance for different concentrations becomes less pronounced and is smaller than the measurement error, i.e., it is not reliable. A realistic output can only be expected for < 20 mM and frequencies of < 10 Hz. We perform fitting of the relationship between impedance magnitude and analyte concentration with either a non-linear phenomenological fit over the entire dynamic range (Fig. 3e dashed lines and Fig. S10) or linear with two linear ranges (0-1 mM and 10-50 mM, see Fig. S11), albeit with significantly reduced fitting quality. A detailed discussion of the fitting is given in SI Section 5.2 and summarized in Table S1.

The advantages that this sensing mode provides are: (i): no impedance data fitting is necessary to extract the sensing information, which allows a very facile readout. (ii): The concentration information can be extracted even at a single frequency < 10 Hz. (iii): Measurement time and sensitivity can be tuned by choosing an adequate measurement frequency. (iv): In principle, the sensitivity range can again be tuned by the illumination time and light intensity and also time delayed readout is possible. (v): The readout is non-invasive and does not alter the concentration information.

*Coulometric sensing* quantifies the concentration of the analyte by discharging the sensor *after* charging it for a fixed time (here 60 s) after illuminating under OCP conditions in a degassed electrolyte containing a fixed glucose concentration. This readout should be distinguished from the others, as it is invasive, i.e., it modifies the concentration information of the sensor during discharge. The discharge is performed by applying a constant potential of +0.2 V vs. Ag/AgCl and measuring the dark current. We observe an initial large current of up to 153 µA, which rapidly decays and plateaus after approximately 60 s (Fig. 3f), at which already 96 % of the accumulated charges are discharged. This fast decline can be explained with a decreasing amount of charge carriers available on the material and hence increasing resistance when discharging, a typical phenomenon for batteries,[52] and in-line with the discussed correlation between resistance (i.e., magnitude of impedance) and amount of charging above for impedimetric sensing. After 300 s of discharge, the average current for all experiments (irrespective of analyte concentration) was below 10 nA, which indicates a near complete discharging process. By integrating the current over the entire 300 s, we obtain a precise measure of the total charge that was stored on the system. This charge contains the sensing information (Fig. 3g), as it mirrors the amount of charging and thus, glucose concentration which had led to charge accumulation on K-PHI at a fixed illumination time as discussed above for potentiometric sensing (Fig. 3b). Note that this experiment also acts as a reset of the sensor for all memristive cases so that it can afterwards be reused for the next concentration sensing experiment, i.e., it deletes all previously stored information. This underlines that the change in memristive state is only dependent on the charging state and fully reversible, an important characteristic of a memristor.[53]

Combining this sensing method with a possible long charge retention time (when illuminating for 20 min, more than 1 h is possible; see Fig. 1d) allows us to demonstrate the delayed sensor readout, i.e., 'memory' sensing. When charging the sensor for 60 s and performing the discharge after delays of 60 and 300 s, a decrease in charge is observed when increasing the delay time (see SI Section 6.1 for details). We attribute this to self-discharge via uncovered parts of the FTO substrate due to charge storage more negative than the thermodynamic potential required for water reduction from the aqueous electrolyte. This self-discharge is also responsible for the larger charge error at low concentrations (Fig. 3g), making this readout more useful for large concentrations. Fitting of the measurement to extract the concentration information is once again done with a phenomenological nonlinear fit over the entire concentration range (Fig. 3g dashed line and Fig. S12) or by a linear fit that requires two linear ranges (0.1-5 mM and 10-50 mM, see Fig. S13) and has a lower fit quality. The delay times produce a systematic feature by either changing fit contributions or an offset for nonlinear and linear fitting, respectively. An offset factor can thus be calculated to account for the charge loss. Details of the fitting are given in SI Section 5.3 and Table S1.

Advantages that the coulometric sensing provide are as follows: The sensitivity can again be tuned by deliberately setting the illumination time and/or intensity during charging: short illuminations are beneficial for large concentration and long illuminations for low concentrations. Compared to the other electrochemical readout methods discussed above, the standard deviation between samples is larger (see Fig. 3g). We attribute this to the differently pronounced self-discharging kinetics and the integrating mechanism, in which small differences among samples such as coating degree and density play a much more significant role. However, since sensor fabrication can be scaled easily (dip coating) and sample batches can be calibrated with an offset factor, these effects are of little practical concern (see SI Section 4).

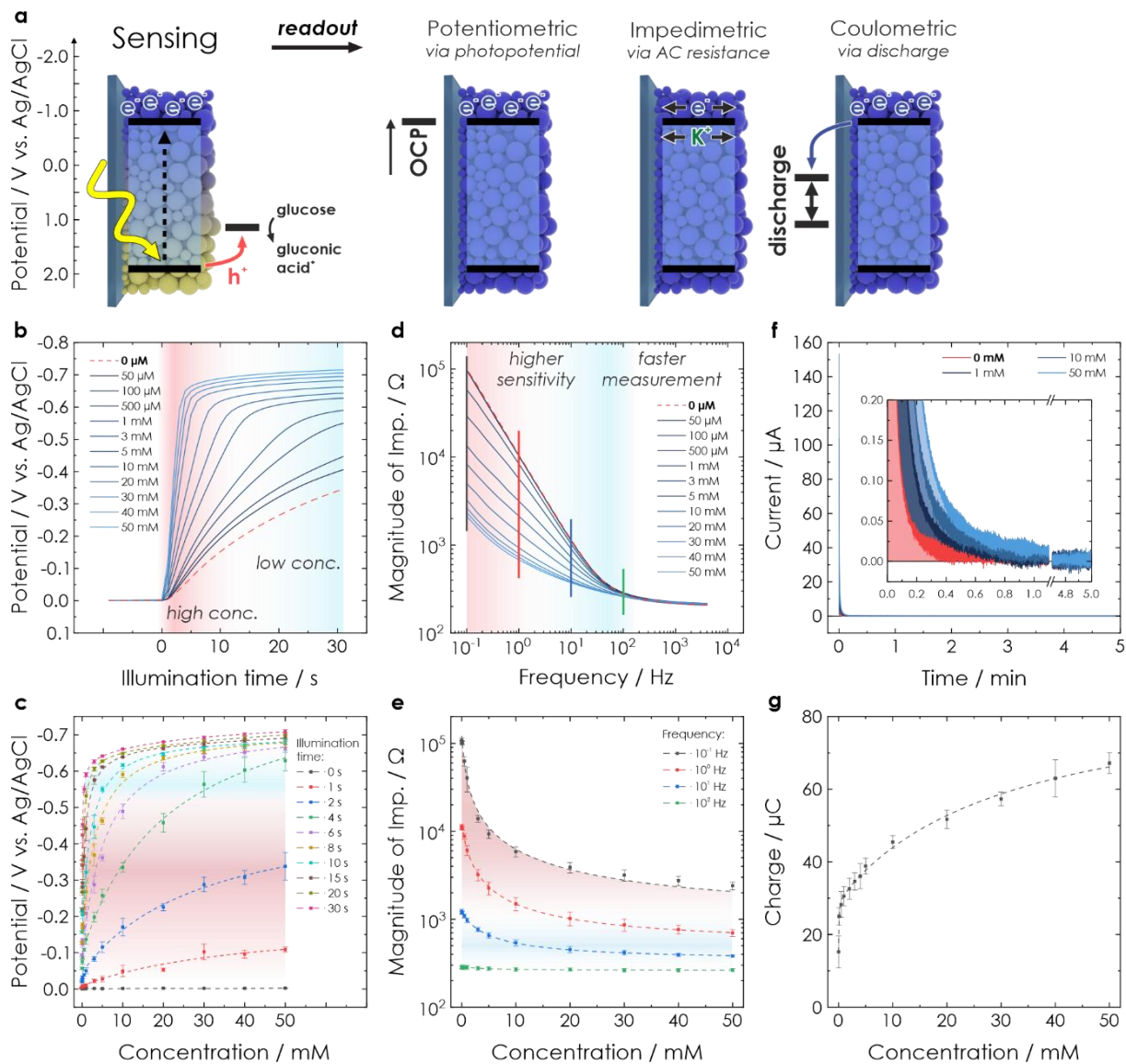

**Fig. 3 | Electrochemical readout methods for the thin-film based memristive K-PHI sensor. a,** Scheme of the sensor and readout methods. First, the sensing information is written into the material (a, left). Subsequently, the charged state can be analyzed via different electrochemical methods (a, right). **b-c,** Potentiometric readout. The OCP of the charged state under illumination increases over time for a given concentration. With increasing concentration of glucose, the OCP shift becomes more pronounced due to faster charging. The concentration of glucose can hence be extracted at different illumination times independently, which allows to tune the sensitivity range. **d-e,** Impedimetric readout. Bode plot of the impedance response after illumination of the sensor for 60 s at OCP condition (see b). A change in magnitude can be correlated to the analyte concentration due to conductivity changes upon memristive charging. This effect has a varying signal amplitude at different time scales (i.e., frequencies; shown in d with vertical lines in d) and concentrations. **f-g,** Coulometric readout after sensing interaction (60 s of illumination at OCP conditions). The charged state is discharged for 300 s at a bias

of +0.2 V vs. Ag/AgCl (f, inset shows a zoom of low currents). The charge correlates monotonously with the glucose concentration. Dashed lines in c, e and g are non-linear fits of the sensor response over the entire dynamic. Fitting is discussed in SI Section 5.1, 5.2 and 5.3.

**Wireless memristive direct sensing with optical readout**

Last, we study the optical response of K-PHI particles suspended in aqueous solution. This sensing mode is more direct and its implementation is potentially simpler compared to the previous ones, as no wires or electrode substrates are needed for *operando* probing. The memristive charge accumulation is caused by the same light-induced oxidative charging process and reflects the accumulated sensing information via a change in photophysical properties (Fig. 4a).

*Colorimetric sensing:* The most prominent change in the photophysical properties of K-PHI due to photocharging is a color change from yellow to blue, the so-called 'blue' state.[24,26,39] Measuring the K-PHI absorption after the sensing interaction allows to quantitatively relate the optical properties to the analyte concentration (Fig. 4b) in suspensions with a given concentration of K-PHI particles (3 mg/ml). After photocharging for 200 s in the absence of oxygen and in the presence of the analyte (here glucose), the absorptance of the suspension was measured from 800 to 350 nm (Fig. 4b). Note that for higher accuracy, absorptance and not absorbance is measured in an integrating sphere, to correct for scattering. With increasing donor interaction, an additional absorption band with a maximum at ~670 nm appears and grows, which is characteristic for the 'blue' state.[24] Control experiments without glucose confirmed no absorptance at wavelengths above the optical bandgap of K-PHI (~ 450 nm). Plotting the value of the absorptance at this maximum against the analyte concentration reveals the colorimetric sensing ability (Fig. 4c). An increase in absorptance is detectable down to a glucose concentration of 200 µM for the chosen illumination time and K-PHI concentration. Similar to the above-discussed electrochemical readout methods, illumination time can be used to tune the response towards a more specific sensitivity range (see SI Section 7.3). Furthermore, a time-delayed readout of the sensing information is easily possible, with a signal decay of 31.8 % after a delay of 20 min (see SI Section 7.4). This enables for example in-situ measurements with ex-situ readout in environments where in-situ optical sensing is not possible. Note that the color change is also clearly visible by eye (Fig. 4c, inset), which makes this readout method useful for label-free colorimetric applications and qualitative analysis even without instrumentation. In comparison to previously reported colorimetric carbon nitride sensors that rely on color changes of external species such as 3,3',5,5'-tetramethylbenzidine (TMB), our approach does not require any additional external signal molecules to achieve a visible color change, minimizing fabrication cost and time while simultaneously improving the design simplicity and recyclability.[33] The sensing information can be erased by opening the cuvette and enabling quenching of the reduced state by oxygen contained in air.[28,30] Notably, the quenched state always restores the initial absorptance value (Fig. 4c red dots) and no degradation of the material's optical properties was observed. Hence, this resetting method is more practical than washing the sensor to reset the concentration of external species (such as TMB). The relationships between absorptance signal and concentration can be fit analogous to the above discussed electrochemical readout methods. We find three linear (0.2 to 2 mM, 3 to 10 mM, 15 to 50 mM, see Fig. S15) ranges. Phenomenological non-linear fitting works over the entire dynamic concentration range (Fig. 4c dashed line and Fig. S14) and is therefore better suited. A more detailed discussion of the fitting is given in SI Section 5.4 and summarized in Table S1.

*Fluorometric sensing:* When exciting the characteristic and broad fluorescence emission of K-PHI at 370 nm, the emission signal with a maximum at ~450 nm can also be used to characterize the charged state, analogous to the colorimetric measurements. After illuminating the K-PHI suspension with different glucose concentrations for 200 s, we measured the emission spectrum from 400 to 600 nm (Fig. 4d). With linearly increasing glucose concentrations, we observed a fast decreasing fluorescence activity (Fig. 4e). Since the absorption at the excitation wavelength (370 nm) remains unchanged (Fig. 4b), this decline is purely an emission property, which is caused by the accumulation of electrons, enabled by hole extraction due to glucose oxidation. We attribute the emission quenching to an increased recombination probability of photogenerated holes, which have more recombination partners with increasing amounts of electrons being stored on the material. This de-excitation, induced by accumulated charges, may be either radiative at frequencies out of our detection limits (> 900 nm) or non-radiative. A schematic summary of this model together with a discussion of the negligible invasiveness of these measurements is given in SI Section 7.1 and 7.2. The glucose concentration cannot only be quantified by integrating the emission spectrum (Fig. 4e), but also by an emission intensity measurement at a single wavelength. By that, the measurement time can be significantly decreased (20 s to 0.1 s). The sensitivity of the fluorometric readouts is comparable to the colorimetric readout, i.e., a change in signal can be detected down to 200 µM for the chosen illumination time of 200 s. Emission from the quenched sensor (Fig. 4e, red dots) without washing the material or exchanging the electrolyte reveals a slight decay in the emission of the discharged state, which we attribute to slow particle agglomeration or surface clogging by oxidized donor species.[49] Both can be restored by washing away the analyte and its oxidation products, or by sonicating the material in order to fully restore the properties of the sensor material.[49] Besides, a slight dilution of K-PHI suspension due to the addition of glucose solution might also contribute to the signal decay. The relation between the analyte concentration and the signal can be fit analogous to the colorimetric readout, with three linear ranges (0.2 to 2 mM, 2 to 10 mM, 15 to 50 mM; see Fig. S17) or one non-linear fit over the entire dynamic concentration range (Fig. 4e dashed line and Fig. S16). A detailed discussion of the fitting is given in SI Section 5.5 and Table S1. Last and similar to before, the sensitivity ranges can be tuned by varying the illumination time (Fig. S18) or intensity and also by increasing the power of the laser for PL excitation. Besides, also time-delayed readout is possible with a signal loss of only 21.1 % after a delay of 1200 s (see SI Section 7.4). In principle, the wireless methods described here are applicable also to K-PHI particles immobilized on a thin film. However, the application of wireless K-PHI particles for sensing holds more promise for applications where environmental or biological conditions are studied *in-situ*.[54]

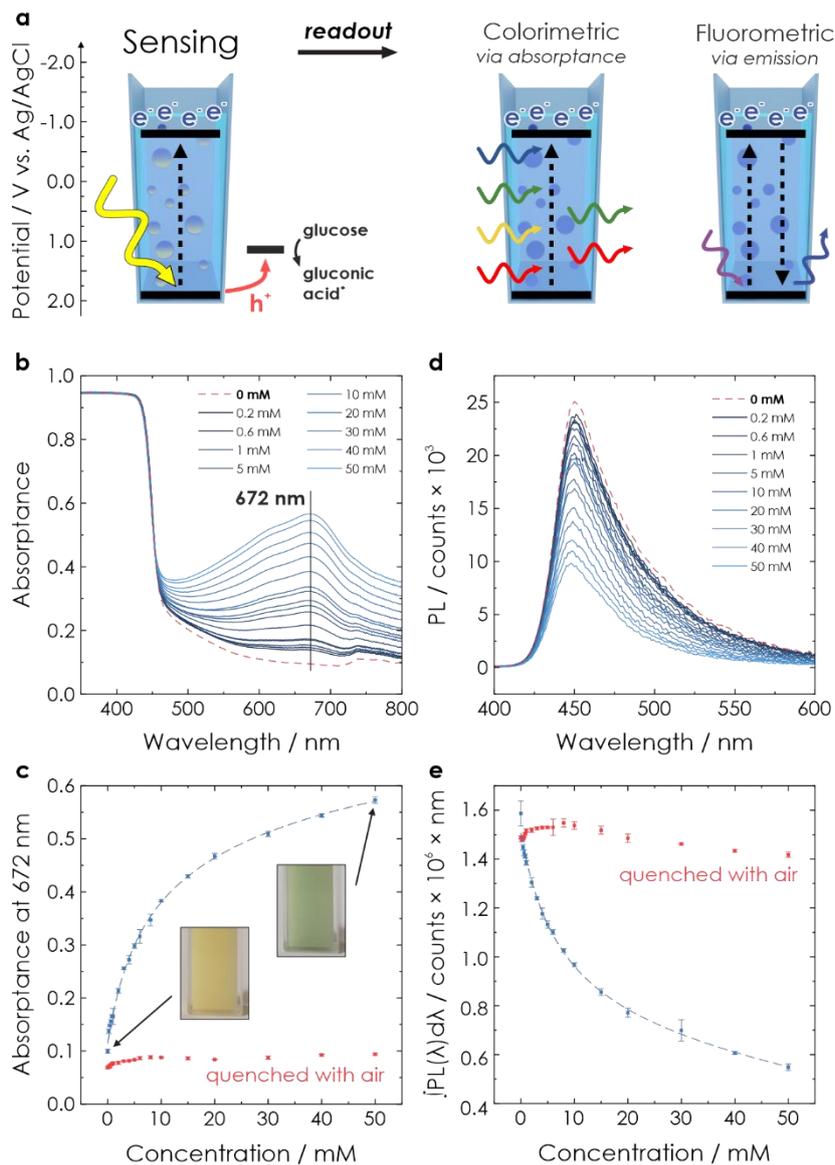

**Fig. 4 | Wireless optical readout of the K-PHI particles acting as direct memristive sensor in aqueous suspension. a,** Scheme of how optical properties can be accessed. **b-c,** Charging is accompanied by a color change of the material (inset pictures in c) and can be quantified via absorptance measurement. **d-e,** Fluorometric readout. When exciting the system at a wavelength of 370 nm after the sensing interaction, the PL emission quenching caused by accumulated charges is quantified and correlated to the glucose concentration. Dashed lines in c and e are non-linear fits of the sensor response over the entire dynamic range. Fitting is discussed in SI Section 5.4 and 5.5.

**Summary and Conclusion**

We have shown that the biocompatible, environmentally friendly K-PHI can be employed for organic analyte sensing in various ways: (i) on a substrate via a PEC amperometric sensing approach, in which an oxidative photocurrent can be correlated to the concentration for different organic and biologically relevant analytes; (ii) *via* a novel memristive sensing concept, which utilizes the material's inherent bifunctionality of light absorption and charge storage to photo-oxidatively sense and quantify analytes by accumulating the generated electrons on the material. The change in the material's optoelectronic properties upon photocharging can be used for various electrochemical and optical readout methods: potentiometric and impedimetric sensing (non-invasive), coulometric sensing (invasive and method to reset), as well as colorimetric and fluorometric sensing (minimal invasive). The latter two optical methods enable wireless analyte quantification for *in-situ* applications as particulate systems and a facile qualitative visual readout.[30,54,55] Contrary to most other sensors, the sensing methods are all direct, i.e., we combine receptor, transducer and current integrating memristive amplifier on the same material, while no additional intermediate species (e.g., mediators) are required. In comparison to state-of-the-art memristive sensors requiring tailored nanostructures, the K-PHI sensor is very simple (only a degassed standard electrochemical cell required), flexible (film electrodes or wireless sensing of particles) and easily scalable (cheap and easy material synthesis, film preparation via solution process). Moreover, the sensor can be reset and reused easily. While linear fitting of the sensor response is possible, non-linear fitting easily allows readout of the entire dynamic range of all sensing methods with the same model. Input parameters for sensing, like illumination time and intensity, can be used to tune the sensitivity and measuring ranges. A time-delayed readout is possible. The main practical advantage of our new sensing approach is its great flexibility for different applications, which enable fast (<1 s) and comparably facile readout, as well as the combination of different techniques for adapting sensitivity ranges without modification of the sensor. A summary of the key advantages and measurement properties of the different analysis methods is given in Table 1. The organic nature of K-PHI with functionalizable groups at its edges was further shown to enable a targeted surface property engineering by chemical modifications, which can lead to much enhanced selectivity.[49] Similarly, the chemical attachment of dedicated receptors appears possible,[56] which would allow to engineer selectivity and responsivity further, according to the field of application. Besides sensing, the concepts described here are useful for evaluating sacrificial electron donor strengths and thereby, they provide a better understanding of photocatalytic processes parameters, while also being of interest as a characterization tool for light driven charge storage properties of 'solar battery' materials. At the same time, these novel sensing approaches can be applied to other photocatalytic and especially light-storing materials, including both organic and inorganic semiconductors. Since the memristive sensor generates an electrical signal upon charging (OCP), which can be used to drive a current by discharging (coulometric measurements), it can be coupled to other (bio)electronic devices in feedback loops, which can be triggered by using voltage or current thresholds. With this, true neuromorphic applications that facilitate automated electronic signal processing would be directly enabled.[5,6,11,57,58]

**Table 1 | Summary of readout methods.** A summary of recommended operation modes as well as key advantages that our new memristive sensing concept provides in comparison to PEC amperometric sensing. Note that all memristive methods are tunable via illumination time (as demonstrated for potentiometric) or illumination intensity. Furthermore, measurement methods in the same bold boxes can be combined, allowing to combine the advantages, minimize the errors and improve accuracy.

| Method | Useful range (glucose sensitivity in the reported conditions) | Illumination time | Readout time | Special characteristics |
|---|---|---|---|---|
| PEC Amperometric | 0.1 mM to 10 mM (11 µM) | continuous | 60 s | - direct sensing<br>- versatile application |
| Potentiometric | 0.05 mM to 50 mM (50 µM) | < 1 mM: **> 10 s**<br>1-10 mM: **4-10 s**<br>> 10 mM: **< 4 s** | instant | - non invasive<br>- memory sensing |
| Impedimetric | 0.05 mM to 20 mM (50 µM) | only 30 s illumination analyzed | < 10 s to 100 s | - non invasive<br>- memory sensing<br>- no fitting required |
| Coulometric | 5 mM to 50 mM (100 µM) | only 30 s illumination analyzed | < 300 s | - memory sensing<br>- sensor reset (invasive) |
| Colorimetric | 0.2 mM to 50 mM (200 µM) | only 200 s illumination analyzed | < 1 s | - minimum invasive<br>- memory sensing<br>- wireless sensing<br>- visible color change |
| Fluorometric | 0.2 mM to 50 mM (200 µM) | only 200 s illumination analyzed | < 1 s | - non invasive<br>- memory sensing<br>- wireless sensing |

**Acknowledgements**

The authors acknowledge Julia Kröger for insightful discussions. We thank Liang Yao for SEM imaging and Alexander Kisser for assistance in electrode preparation. Financial support is gratefully acknowledged by the Max Planck Society, the European Research Council (ERC) under the European Union's Horizon 2020 Research and Innovation Program (grant agreement no. 639233, COFLeaf), the Deutsche Forschungsgemeinschaft (DFG) via the cluster of excellence 'e-conversion' (project number EXC2089/1–390776260) and by the Center for NanoScience (CENS). A.J.S. gratefully acknowledges a postdoctoral scholarship from the Max Planck Society.

**Authors Contributions**

AG, AJS, FP and BVL conceived the project. AG performed the electrochemical measurements. AG and AJS performed the optical measurements. AG and AJS, with assistance of FP, analyzed the data. AJS, NVB, FP and BL supervised the research. AG, AJS and FP wrote the manuscript with assistance of all authors.

**Conflict of interest**

The authors declare no conflict of interest.

**Methods**

**Synthesis of the carbon nitride modification K-PHI as nanoparticles.** The carbon nitride materials were synthesized as described in literature.[24,26,35] The precursor material melamine as well as potassium thiocyanate were acquired from *Sigma Aldrich* in reagent grade purity. Exfoliation was carried out in 2-propanol (IPA) via sonication in an ice bath for 2 h (300 mg K-PHI in 100 mL IPA). Subsequently the nanosheets were separated via two centrifugation steps at 353 RCF for 20 min and 795 RCF for 40 min in a centrifuge (*3-30k, Sigma*) to ensure a uniform small particle size, akin to a reported procedure.[28] To reach the desired concentration, density was first evaluated by drying 1 mL of suspension and measuring the weight of the dried residue on a quartz crystal microbalance. To increase the particle concentration to 0.2 mg/mL excess IPA was then removed using a rotary evaporator (*Hei-Vap Value Digital G3B, Heidolph*) at a pressure of 137 mbar and a water bath temperature of 50 °C.

**Material's Characterization.** ATR-IR spectra of K-PHI bulk suspensions and nanosheets were collected with an IR spectrometer (*UATR TWO, PerkinElmer*), which was equipped with a diamond crystal. The optical bandgap of K-PHI bulk suspensions and nanosheets was characterized with an UV-VIS spectrometer (*Cary 5000, Agilent*), equipped with an integrating sphere. AFM was performed with an AFM microscope (*MFP-3D, Asylum Research*).

**Preparation of the sensor films and suspensions.** Thin films of K-PHI nanoparticles were deposited onto FTO substrates (*Sigma Aldrich*, surface resistivity of 7 $\Omega/cm^2$) via dip coating with 400 dips, 100 mm/min extraction speed and 120 s drying time at ambient temperature between the dips (*ND-R Rotary Dip Coater, Nadetech*). A subsequent annealing at 70 °C for 2 days was performed to ensure removal of all leftover solvent. The sample was then cut into 10x12 mm and a small part of the film was scratched off for contacting. This was performed by gluing a wire to an uncovered part of the FTO using conductive silver paste (*Silver Conductive RS 186-3600, RS-Pro*). The contact was then sealed with epoxy glue (*DP410, 3 M Scotch-Weld*) to provide a rigid connection and prevent both the silver paste and uncovered FTO to influence the measurements. Last, an electrochemical self-cleaning of the finished samples was performed according to a procedure described in SI Section 3.

For measurements, which utilize particles in a suspension, dried as synthesized bulk K-PHI was suspended in water (3 mg/mL) and vortexed for 120 s to ensure a proper distribution of particles.

**Electrochemical measurements.** All electrochemical measurements were performed in an aqueous electrolyte which contained a 10 mM KCl (*Sigma Aldrich*) background electrolyte. The electrolyte was purged with >99 % argon for at least 20 min through a porous glass frit before every measurement. An oxygen content of <10 ppb during measurements was ensured by measuring trace oxygen with a trace optical oxygen meter (*PSt6 sensor spot and Fibox 3 trace, Presens*). All analytes (*Sigma*

*Aldrich*) were dissolved in deionized (DI) water and added to the electrolyte in respective concentrations. An Ag/AgCl electrode with saturated KCl electrolyte (*RE-1CP, ALS Japan*) was used as the reference electrode and a gold foil (*Sigma Aldrich*) as counter electrode. Measurements were carried out with a multichannel potentiostat (*Autolab M204, Metrohm*) in a glass reactor equipped with a Quartz window for illumination.[28] Impedance measurements were carried out with a single-channel potentiostat (*CompactStat, Ivium*). Impedance fitting was performed with the *RelaxIS 3* software, *rhd Instruments*.

Illumination (1 sun) was provided by a calibrated *Sciencetech LightLine A4* solar simulator, which provides simulated sunlight with class AAA quality (AM 1.5G). Light was turned on and off using a *ThorLabs SHB1T* shutter.

**Optical measurements.** All optical measurements were performed in a quartz cuvette (*Hellma Analytics*). The suspension (K-PHI in DI water, 3 mg/mL) was purged with >99 % argon before every measurement for 300 s and during charging illumination. Glucose (*Sigma Aldrich*) was dissolved in DI water and added into the cuvette in respective concentrations. The sample was illuminated (AM1.5G) using a solar simulator (*IEC/JIS/ASTM, Newport*). Fluorescence was measured using a spectrofluorometer (*FLS980, Edinburgh Instruments*). Absorptance spectra were measured using a spectrophotometer equipped with an integrating sphere (*Cary 5000 UV-VIS, Agilent*). The sample was positioned in the center of the sphere on an angle to obtain both total transmission and total reflection signals.